\documentclass[floats,floatfix,showpacs,amssymb,prd,twocolumn,superscriptaddress,nofootinbib]{revtex4-1}

\usepackage{amssymb,amsmath,verbatim,mathtools,needspace,enumitem,etoolbox,graphicx,physics,microtype}

\usepackage[dvipsnames, usenames]{xcolor}

\definecolor{linkcolor}{rgb}{0.0,0.3,0.5}
\usepackage[unicode, colorlinks=true, linkcolor=linkcolor, citecolor=linkcolor, filecolor=linkcolor,urlcolor=linkcolor, pdfusetitle]{hyperref}
\usepackage[all]{hypcap}
\usepackage[T1]{fontenc}
\usepackage[utf8]{inputenc}

\usepackage[utf8]{inputenc}
\usepackage{graphicx,amssymb,amsmath,amsthm,amsfonts,epsfig,times,natbib}

\usepackage{epstopdf}
\usepackage{tensor}
\usepackage{amsbsy}
\usepackage{bm,url}
\usepackage{float}
\usepackage{dcolumn}
\usepackage{latexsym}
\usepackage{rotating}
\usepackage{longtable}
\usepackage{enumerate}
\usepackage{booktabs}
\usepackage[utf8]{inputenc}
\definecolor{oxfordblue}{rgb}{0.0, 0.13, 0.28}
\definecolor{burgundy}{rgb}{0.5, 0.0, 0.13}
\definecolor{darkolivegreen}{rgb}{0.33, 0.42, 0.18}
\definecolor{darkblue}{rgb}{0,0,0.5}
\definecolor{richcarmine}{rgb}{0.84, 0.0, 0.25}
\definecolor{darkblue}{rgb}{0,0,0.5}
\definecolor{venetianred}{rgb}{0.78, 0.03, 0.08}
\definecolor{skobeloff}{rgb}{0.0, 0.48, 0.45}
\hypersetup{colorlinks=true, citecolor=darkblue, linkcolor=darkblue,
urlcolor = darkblue}

\newcommand{\olemiss}{\affiliation{Department of Physics and Astronomy, The University of Mississippi, University, MS 38677, USA}}
\newcommand{\jhu}{\affiliation{Department of Physics and Astronomy, Johns Hopkins University, 3400 N. Charles
Street, Baltimore, MD 21218, USA}}

\def\nn{\nonumber}

\newcommand{\ben}{\begin{enumerate}}
\newcommand{\een}{\end{enumerate}}

\def\be{\begin{equation}}
\def\ee{\end{equation}}
\def\bea{\begin{eqnarray}}
\def\eea{\end{eqnarray}}
\def\nn{\nonumber}
\newcommand{\beq}{\begin{eqnarray}}
\newcommand{\eeq}{\end{eqnarray}} 
\newcommand{\ba}{\begin{align}}
\newcommand{\ea}{\end{align}}

\def\nn{\nonumber}
\def\be{\begin{equation}}
\def\ee{\end{equation}}
\def\beq{\begin{eqnarray}}
\def\eeq{\end{eqnarray}}
\def\f{\frac}

\def\apjl{\rm{ApJ}}

\newcolumntype{L}{>{$}l<{$}}
\newcolumntype{R}{>{$}r<{$}}
\newcolumntype{C}{>{$}c<{$}}
\begin{document}

\title{Multi-mode black hole spectroscopy}

\author{Vishal Baibhav}
\email{baibhavv@gmail.com}
\jhu \olemiss

\author{Emanuele Berti}
\email{berti@jhu.edu}
\jhu \olemiss

\date{\today}

\begin{abstract}
  The first two LIGO/Virgo observing runs have detected several black hole binary mergers. One of the most exciting prospects of future observing runs is the possibility to identify the remnants of these mergers as Kerr black holes by measuring their (complex) quasinormal mode frequencies. This idea -- similar to the identification of atomic elements through their spectral lines -- is sometimes called ``black hole spectroscopy''. Third-generation Earth-based detectors and the space-based interferometer LISA could measure {\em multiple} spectral lines from different multipolar components of the radiation, and therefore provide qualitatively better tests of the Kerr hypothesis. In this paper we quantify the redshift out to which the various modes would be detectable (or, conversely, the number of detectable modes at any given redshift) as a function of the intrinsic parameters of the merging binary. LISA could detect so many modes that current numerical relativity simulations would not be sufficient to extract all available science from the data.
\end{abstract}

\maketitle
%%%%%%%%%%%%%%%%%%%%%%%%%%%%%%%%%%%%%%%%%%%%
\section{Introduction}
%%%%%%%%%%%%%%%%%%%%%%%%%%%%%%%%%%%%%%%%%%%%
The first detection of black hole (BH) binary mergers by the LIGO/Virgo collaboration, GW150914~\cite{Abbott:2016blz}, marked the beginning of gravitational wave astronomy.  The first two observing runs (O1 and O2) led to the detection of 5 confirmed BH binary mergers, a BH binary merger candidate which is likely to be of astrophysical origin~\cite{TheLIGOScientific:2016pea}, and a neutron star binary~\cite{TheLIGOScientific:2017qsa}. Therefore the inspiral, merger and ringdown of compact objects is anticipated to be the main target of the next LIGO/Virgo observing run (O3).

In this paper we focus on the so-called ``ringdown'' stage of a BH binary merger, where the deformed remnant relaxes to a Kerr BH. Out of the events observed so far, only one (GW150914) had significant signal-to-noise ratio (SNR) in the ringdown, but it is quite likely that O3 will lead to more and louder observable ringdown events. The ringdown is a sum of damped oscillations known as ``quasinormal modes,'' with frequencies and damping times that depend only on the mass and spin of the final BH~\cite{Kokkotas:1999bd,Berti:2009kk,Konoplya:2003ii}. The simplicity of the spectrum allows us to identify a Kerr BH, just like spectral lines can be used to identify atomic elements: this idea is commonly referred to as ``black hole spectroscopy''~\cite{Detweiler:1980gk,Dreyer:2003bv,Berti:2005ys}.  Modifications of general relativity may or may not lead to BH solutions that differ from the Kerr solution~\cite{Berti:2015itd}, but the {\em dynamics} of these solutions and their gravitational-wave emission will, in general, differ from general relativity~\cite{Cardoso:2016ryw,Berti:2018vdi}. The ringdown can be modified even within general relativity if the merger remnant is some exotic compact objects -- such as a boson star -- or if there are significant modifications in BH dynamics at the horizon scale, as suggested by some quantum gravity models~\cite{Cardoso:2017cqb}.

Spectroscopic tests of Kerr dynamics require the measurement of multiple quasinormal mode frequencies~\cite{Dreyer:2003bv,Berti:2005ys,Berti:2007zu,Gossan:2011ha}. The fundamental (and loudest) mode is needed to extract the mass and spin of the remnant. Any other mode can then be used to look for departures from general relativity or constrain their magnitude.  However, the detectability of each quasinormal mode is contingent on whether it is excited to high enough amplitude in the merger.  In general relativity, the specific nature of the perturbation does not affect the quasinormal mode frequencies, but it affects the degree to which different modes are excited~\cite{Leaver:1986gd,Andersson:1995zk,Berti:2006wq,Zhang:2013ksa}. The excitation (and hence the detectability) of different quasinormal modes in a binary BH coalescence depends on the properties of the progenitors in a way that can be quantified using numerical relativity simulations~\cite{Buonanno:2006ui,Berti:2007fi,Berti:2007zu,Kamaretsos:2011um,London:2014cma,Kamaretsos:2012bs,Baibhav:2017jhs,London:2018gaq}.

Significant detector improvements may be necessary to detect ringdown with high SNRs, or to detect sub-dominant modes~\cite{Berti:2016lat,Bhagwat:2016ntk}. The prospects for detecting high-SNR events or multiple modes will be much better with third-generation ground-based detectors -- like the Einstein Telescope (ET)~\cite{Punturo:2010zz} or Cosmic Explorer (CE)~\cite{Dwyer:2014fpa,Evans:2016mbw} -- and with the space interferometer LISA~\cite{AmaroSeoane:2012km,Audley:2017drz,Armano:2018kix}. In the absence of a direct measurement of higher-order modes, spectroscopic tests of general relativity may still be possible with current-generation detectors by combining posterior probability densities from multiple detections~\cite{Meidam:2014jpa} or via coherent stacking~\cite{Yang:2017zxs}.

Astrophysically, BH masses range from $\sim3 M_\odot$ to $10^{10} M_\odot$ (see e.g.~\cite{Barack:2018yly} for a recent review). Prior to the direct detection of gravitational waves, stellar-mass BHs were known to exist in X-ray binaries with masses ranging from $\sim 3 M_\odot$ to $\sim 20~M_\odot$~\cite{Casares:2013tpa}. We now know that stellar collapse can generate BHs as massive as $\sim 36~M_\odot$ (unless the progenitors of LIGO mergers were themselves formed in previous mergers~\cite{Gerosa:2017kvu,Fishbach:2017dwv,Fishbach:2017zga}). Theory extends this range up to $\sim 40$--$60~M_\odot$ and predicts the existence of a ``mass gap'' between $\sim 60$--$150~M_\odot$, because in this mass window pair instabilities during oxygen burning can lead either to substantial mass losses or (in higher-mass stellar progenitors) to the complete disruption of the star~\cite{Belczynski:2016jno}. ``Stellar-mass'' BHs heavier than $150~M_\odot$ can form at low metallicities if the initial mass function of stars extends further out, up to hundreds of solar masses.  There is circumstantial observational evidence for IMBHs: they have been claimed to source ultra-luminous X-Ray sources~\cite{Colbert:1999es}, with further claims of detection in star clusters~\cite{Maccarone:2007dd,Gebhardt:2005cy,Gebhardt:2002in} and from quasi-periodic oscillation~\cite{Caballero-Garcia:2013vxa,Pasham:2015tca}. Second- and third-generation gravitational-wave detectors are sensitive to ringdown from intermediate-mass BHs (IMBH), so ringdown observations can help shed light on the nature and extent of the mass gap and on the existence of IMBHs. LISA~\cite{AmaroSeoane:2012km,Audley:2017drz,Armano:2018kix}, currently scheduled for launch in 2034, will target more massive BH mergers.

Our main goal in this work is to assess the capabilities of these future gravitational-wave detectors to observe multiple ringdown modes. The plan of the paper is as follows. In Section~\ref{sec:detection} we review the criteria to detect multiple ringdown modes and the calculation of their SNR.
In Section~\ref{sec:horizon} we compute the horizons out to which higher-order modes would be detectable, we define and quantify the response redshift and detectability fraction, and we point out interesting features in the response redshift for higher-order modes. We conclude in Section~\ref{sec:conclusions} by pointing out the limitations of this study and directions for future work.

%%%%%%%%%%%%%%%%%%%%%%%%%%%%%%%%%%%%%%%%%%%%%%%%%
%%%%%%%%%%%%%%%%%%%%%%%%%%%%%%%%%%%%%%%%%%%%%%%%
\section{Detectability and signal-to-noise ratio of higher-order ringdown modes}
\label{sec:detection}
%%%%%%%%%%%%%%%%%%%%%%%%%%%%%%%%%%%%%%%%%%%%%%%%%%%%%
%%%%%%%%%%%%%%%%%%%%%%%%%%%%%%%%%%%%%%%%%%%%%%%%%%%%%

In this section we outline a method -- the Generalized Likelihood Ratio Test (GLRT) -- that can be used to test whether a given mode with multipolar indices $(\ell\,,m)$ is present in the ringdown signal. The GLRT was used in~\cite{Berti:2007zu} to study ringdown detectability in the time domain under the simplifying assumption of white noise. In general the noise in a gravitational wave detector is colored, so different ringdown modes for the same merging binary BH system will be affected by noise in a different way. Here we work in the frequency domain and, for simplicity, we assume that all dominant modes (besides the one we are looking for) are known and have been subtracted from the signal. In the same spirit, we also ignore the parameter estimation noise that arises from subtracting imperfectly estimated dominant modes~\cite{Yang:2017zxs}.

Let $n(t)$ be the noise, and $y(t)$ the signal that is left after all dominant modes have been subtracted. Call ${\cal H}_1$ the hypothesis that the signal contains the next subdominant $(\ell\,,m)$ mode, and ${\cal H}_2$ the hypothesis that it does not:
\be
\begin{cases}
{\cal H}_1: y(t) = A\,h_{\ell m}(t)+n(t)\,,\\
{\cal H}_2: y(t) = n(t)\,.
\end{cases}
\ee
The likelihood that the $(\ell\,,m)$ mode (with unknown amplitude $A$) is present is then given by
\begin{equation}
{\rm P}_{A} \propto e^{- \langle y-A h_{\ell m} |y-A h_{\ell m} \rangle}\,,
\end{equation}
where
\begin{align}
 \langle h_1 | h_2\rangle \equiv 2 \int^\infty_0 \frac{\tilde h_1^* \tilde h_2 +\tilde h_1 \tilde h_2^*}{S_h} df.
 \end{align}
By extremizing the likelihood given above, i.e. by computing
\be
%\max_{A} {\rm P}_{A} &=&
{\rm max}_{A} \ln {\rm P}_{A} \nn\\
= {\rm min}_{A} \langle y-A h_{\ell m} |y-A h_{\ell m} \rangle \,,
\ee
we can evaluate the maximum-likelihood estimate $\hat{A}$ of the
unknown parameter $A$, with the result:
\begin{equation}
\hat{A} =  \frac{\langle y |  h_{\ell m} \rangle }{\langle h_{\ell m} | h_{\ell m} \rangle}  \,.
\end{equation}
We now compute the logarithm of the ratio of the maximized likelihoods under the two hypotheses:
 \be
 T(y) = \ln \frac{{\rm max}_{\mathcal{H}_1} P_{A}}{{\rm max}_{\mathcal{H}_2} P_{A=0}}
 = \frac{\hat{A}^2}{2} \langle h_{\ell m} | h_{\ell m} \rangle
 =\frac{\langle y |  h_{\ell m} \rangle ^2}{2\langle h_{\ell m} | h_{\ell m} \rangle}\,.
 \ee
According to the GLRT test, we favor the hypothesis ${\cal H}_1$ if $\sqrt{2T(y)}$ exceeds a specified threshold $\gamma$:
\begin{align}
\label{eq:T_y-threshold}
\sqrt{2 T(y)} =\rho_{\rm crit}=\hat{A}~|| h_{\ell m} || >\gamma\,,
\end{align}
where we have defined $\rho_{\rm crit}^2 \equiv \langle \hat{A} h_{\ell m} | \hat{A} h_{\ell m} \rangle$.

We choose $\gamma$ by setting a tolerable false-positive rate $P_f= Q(\gamma)$,  where
\begin{align}
Q(x) \equiv \frac{1}{\sqrt{2\pi} } \int^\infty_x e^{-\frac{z^2}{2}} dz
\end{align}  
is the right-tail probability function for a Gaussian distribution with zero mean and unit variance.
The detection rate $P_d$ is given by  
\be
P_d =  Q(\gamma-\rho_{\rm crit}) = Q (Q^{-1}(P_f) -\rho_{\rm crit})\,.
\ee
From these criteria we can compute the critical SNR required to claim detection of a given mode:
\be
\rho_{\rm crit}=  Q^{-1}(P_d) - Q^{-1}(P_f) \,.
\ee
For example, by choosing $(P_f \,, P_d)=(10^{-6}\,,0.99)$ we would get $\rho_{\rm crit}=7.08$, close to the threshold of $8$ used by the LIGO Scientific Collaboration. We will follow the LIGO convention and choose a more stringent threshold of $\rho_{\rm crit}= 8$~\cite{Thorne:1987af}.

%%%%%%%%%%%%%%%%%%%%%%%%%%%%%%%%%%%
\subsection{Signal-to-noise ratio}
%%%%%%%%%%%%%%%%%%%%%%%%%%%%%%%%%%%

Ref.~\cite{Berti:2007zu} defines the total SNR in the time domain as $\rho=||\sum_{\ell m} h_{\ell m}(t)||/\sigma$, where $\sigma$ is the assumed {\em white} noise. Since $\sigma$ is proportional to the norm of the sub-dominant mode, the simplifying assumptions made in ~\cite{Berti:2007zu} lead to a detector-independent criterion for mode detectability.

Here we work under the more realistic assumption that the noise is colored and we integrate over the noise power spectral density of the detector $S_h(f)$, so we do not follow the procedure of~\cite{Berti:2007zu} to compute the total SNR. Instead we compute the ringdown SNR from a BH of mass $M$ at distance $r$ as
 \be \label{eq:defSNR}
\rho^2 = 4\int_0^\infty \f{\tilde h^*(f) \tilde h(f)}{S_h(f)}df\,,
\ee
where $\tilde h(f)$ is the Fourier transform of the gravitational wave strain.  Focusing on the fundamental ($n=0$) mode for a given multipolar component $(\ell,\,m)$, the two ringdown polarizations after summing over the $+m$ and $-m$ modes are given by~\cite{Berti:2007zu,Kamaretsos:2011um,Kamaretsos:2012bs}
\begin{eqnarray}
h_{+}^{\ell m}(t) & = & \f{M {\cal A}_{\ell m} Y_{+}^{\ell m}}{r}\;
\Re(e^{-t/\tau_{\ell m}+i(2 \pi f_{\ell m} t+\phi_{\ell m})}) \,, \nonumber \\
h_\times^{\ell m}(t) & = & \f{M {\cal A}_{\ell m} Y_\times^{\ell m}}{r}\;
\Im(e^{-t/\tau_{\ell m}+i(2 \pi f_{\ell m} t+\phi_{\ell m})}) \,, 
\end{eqnarray}
where $f_{\ell m}$ is the quasinormal frequency, $Q_{\ell m}$ is the quality factor, $\tau_{\ell m}=Q_{\ell m}/(\pi f_{\ell m} )$ is the damping time, and the angular functions are defined as
\begin{eqnarray}
Y^{\ell m}_+(\iota) & \equiv & {_{-2}Y^{\ell m}}(\iota,0) + (-1)^\ell\,  {_{-2}Y^{\ell -m}}(\iota,0),\nonumber\\
Y^{\ell m}_\times(\iota) & \equiv & {_{-2}Y^{\ell m}}(\iota,0) - (-1)^\ell\,  {_{-2}Y^{\ell -m}}(\iota,0).
\end{eqnarray}

%%%%%%%%%%%% Plot Full Horizon%%%%%%%%%%%%%%%%%%%
\begin{figure*}[htp]
  \includegraphics[width=\textwidth]{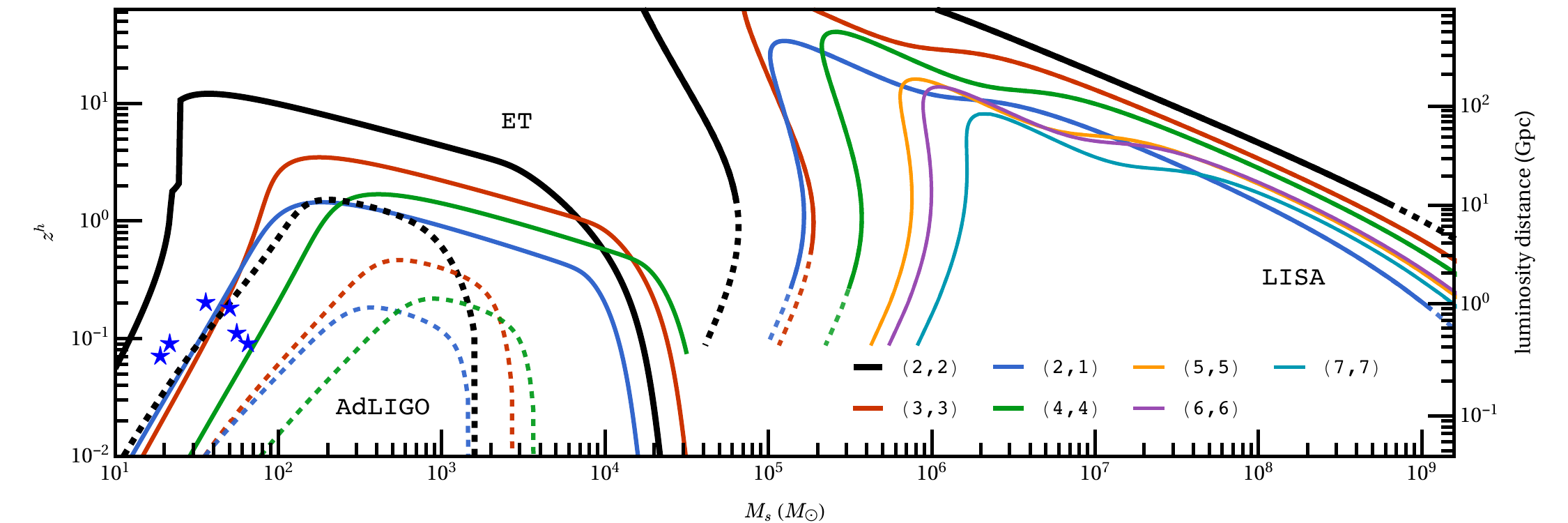}\\ 
  \includegraphics[width=\textwidth]{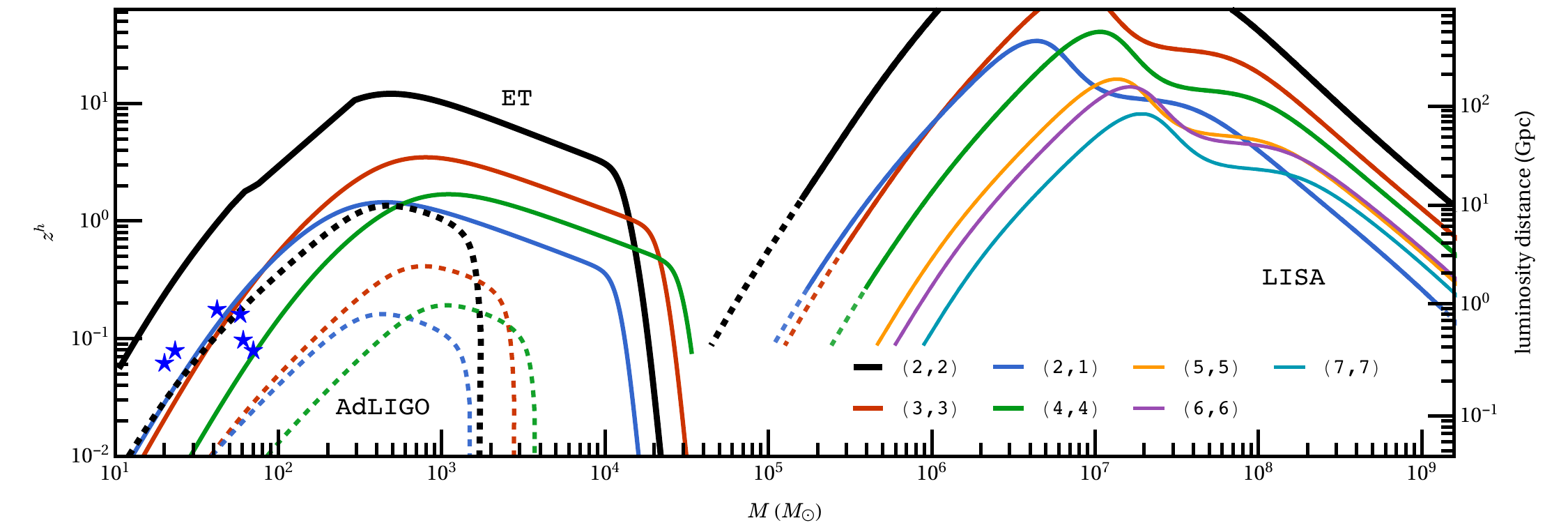}
  \caption{Horizon redshift (left scale) and luminosity distance (right scale) as a function of the remnant BH mass in the source frame (top panel) and in the detector frame (bottom panel) for an optimally oriented, nonspinning BH binary merger with mass ratio $q=2$ as observed by ET (solid lines), Advanced LIGO (dashed lines) and LISA. Star symbols ($\star$) mark the mass and redshift of the six binary BHs detected by the LIGO/Virgo collaboration so far.} 
 \label{fig:fullHorizon}
\end{figure*}
%%%%%%%%%%%%%%%%%%%%%%%%%%%%%%%%%%%%%%%%%%%%%%%%%%%%%%%

The strain measured by the detector is
\be
h = h_{+} F_+ + h_{\times}F_\times \,,
\ee
where  $F_{+,\times}$ denotes the pattern functions (see e.g.~\cite{Cutler:1994ys}):
%depends on pattern functions,
\bea
F_+ &=& \frac{1}{2}(1 + u^2) \cos 2\phi \cos2\psi-u \sin 2\phi \sin 2\psi\;, \nn \\
F_{\times} &=& \frac{1}{2}(1 + u^2 ) \cos2\phi \sin 2\psi
+ u \sin 2\phi \cos 2\psi\;.
\label{pattern}
\eea
Here we use the standard notation for the angles $(\theta,\,\phi)$ describing the source location in the sky and for the polarization angle $\psi$, and we define $u=\cos \theta$.
To compute the SNR, we follow Flanagan and Hughes~\cite{Flanagan:1997sx}: we assume that the waveform for $t<0$ is identical to the waveform for $t>0$, and we divide the amplitude by  $\sqrt{2}$ to compensate for the doubling. Proceeding as in~\cite{Berti:2005ys} we find that
\be \label{eq:snrShort}
\rho_{\ell m}^2=
\left(\f{M {\cal A}_{\ell m} \Omega_{\ell m}}{r}\right)^2  \f{\tau_{\ell m}}{2 S_h(f_{\ell m})}\,,
\ee
where we have defined the sky sensitivity for the given multipole as $\Omega_{\ell m}\equiv \sqrt{\left(F_+ Y ^{\ell m}_+\right)^2+\left(F_\times Y ^{\ell m}_\times\right)^2}$ (see e.g.~\cite{Dominik:2014yma,Chen:2017wpg}).
The quasinormal mode amplitude $ {\mathcal A}_{\ell m}$  is related to the radiation efficiency $\epsilon_{\rm rd}\equiv E_{\ell m}/M $ through~\cite{Flanagan:1997sx,Berti:2005ys}
\be
{\mathcal A}_{\ell m}=\sqrt{\frac{4 \epsilon_{\rm rd}}{M Q_{\ell m} f_{\ell m}}}\,.
\ee
To calculate the radiated energy $E_{\ell m}$ we use fits of form~\cite{Baibhav:2017jhs}
\be
E_{\ell m}=
\left[a_{\ell m}(q)+b_{\ell m}(q) \,\chi_+ + c_{\ell m}(q) \chi_-\right]^2\,.
\label{eq:fit}
\ee
Here $a_{\ell m}$, $b_{\ell m}$ and $c_{\ell m}$ are functions of the binary's mass ratio $q=m_1/m_2\geq 1$ and of the effective spin parameters $\chi_{\pm}$, which in turn are defined in terms of the masses $(m_1,\,m_2)$ and dimensionless spins $(\chi_1,\,\chi_2)$ of the merging BHs as
\be
\chi_{\pm}\equiv \frac{m_1 \chi_1\pm m_2 \chi_2}{m_1+m_2}\,.
\ee
In particular, $\chi_+$ (sometimes denoted as $\chi_{\rm eff}$) is the total ``effective spin'' parameter measured by LIGO, which is conserved at second post-Newtonian order during the binary's evolution~\cite{Racine:2008qv,Kesden:2010yp,Kesden:2014sla,Gerosa:2015tea,Gerosa:2015hba}.  To leading order, the excitation of the $(2\,,1)$ mode depends solely on the other (``asymmetric'') spin parameter $\chi_-$, with a functional dependence of the form $E_{21}= \left[f(q)+g(q)\chi_-\right]^2$ (see~\cite{Berti:2007nw} for details).

%%%%%%%%%%%%%%%%%%%%%%%%%%%%%%%%%%%%%%%%%%%%%%%%%%%%%%%%%
\section{Horizon redshift, response redshift and detectability fraction}
\label{sec:horizon}
%%%%%%%%%%%%%%%%%%%%%%%%%%%%%%%%%%%%%%%%%%%%%%%%%%%%%%%%%%%% 
We can rewrite the SNR in Eq.~(\ref{eq:snrShort}) as
\be
\rho_{\ell m}=\rho_{\rm opt} w_{\ell m}\,,
\ee
where
\be \label{eq:snrOPt}
\rho_{\rm opt}^2=
\left(\f{M {\cal A}_{\ell m} \Omega_{\ell m}^{\rm max}}{r}\right)^2  \f{\tau_{\ell m}}{2 S_h(f_{\ell m})}
\ee
is the SNR for a binary that is optimally located and oriented in the sky,
and $w_{\ell m}(\theta,\phi,\psi,\iota)\equiv \Omega_{\ell m}/\Omega_{\ell m}^{\max}$
is a ``projection function'' such that $0 \le w_{\ell m}\le 1$.
We define the ``horizon redshift'' $z^h$ and the corresponding horizon luminosity distance $d_{L}^h$ (computed using the standard cosmological parameters determined by \emph{Planck}~\cite{Ade:2015xua}) as the farthest distance (or redshift) at which the ringdown from a given mode can be detected, or -- according to our conventions -- the redshift at which $\rho_{\rm opt}(z^{h})=8$. Note that the notion of ``optimally-oriented" has a different meaning for different modes.

Figure~\ref{fig:fullHorizon} shows the detector horizons as a function of the source-frame remnant mass $M_{\rm s}$ (top) and of the detector-frame remnant mass $M=M_{\rm s}(1+z)$ (bottom) for nonspinning binaries with mass-ratio $q=2$. Stars indicate the mass and redshift of the six LIGO detection candidates so far (including the astrophysical candidate LVT151012~\cite{TheLIGOScientific:2016pea}). Advanced LIGO at design sensitivity could detect the dominant $(2\,,2)$ mode from a $\approx 60 M_\odot$ GW150914-like binary out to redshifts $z\simeq 0.36$, but the horizon redshift would be sensibly larger ($z\simeq 0.87$) for the merger of two $\sim 50 M_\odot$ mass BHs, if such massive BHs are indeed formed by either stellar collapse or repeated mergers~\cite{Belczynski:2016jno,Gerosa:2017kvu,Fishbach:2017dwv,Fishbach:2017zga} .

Significant improvements over current detectors are necessary to detect sub-dominant modes from BH binary mergers similar to those observed so far. Therefore, for the time being, we must resort to combining posterior probability densities from multiple detections~\cite{Meidam:2014jpa}, coherent stacking~\cite{Yang:2017zxs}, or possible narrow-band tuning~\cite{Tso:2018pdv} to boost the detectors’ sensitivity in order to test general relativity. The situation is drastically different for third-generation detectors like the Einstein Telescope\footnote{In this paper we use the ET-B noise
  power spectral density available at \href{http://www.et-gw.eu/index.php/etsensitivities}{http://www.et-gw.eu/index.php/etsensitivities}.} (ET)~\cite{Punturo:2010zz} and
Cosmic Explorer (CE)~\cite{Dwyer:2014fpa,Evans:2016mbw}.
For ET, the dominant $(2\,,2)$ mode would be detectable out to redshift $z\sim 15$ for optimally oriented binaries. Moreover, for a GW150914-like binary, ET could observe the $(3\,,3)$ and $(2\,,1)$ modes out to $z\sim0.1$. Higher-order modes are more excited when the mass ratio is significantly different from unity~\cite{Berti:2007fi}: for example, Fig.~\ref{fig:fullHorizon} shows that the $(3\,,3)$ mode is detectable out to $z\sim 3$ when $q=2$.

One feature of Fig.~\ref{fig:fullHorizon} is noteworthy and requires some explanation. It has long been known that, in the eikonal limit, quasinormal modes can be understood as perturbations of null rays at the light ring that slowly leak out to infinity~\cite{Press:1971wr,1972ApJ...172L..95G,Schutz:1985zz,Cardoso:2008bp}. This interpretation leads to the conclusion that the real part of the frequency of modes with $\ell=m$ scales like $\ell$. Comparisons with numerical results show that this scaling is surprisingly good also for low $\ell$'s~\cite{Ferrari:1984zz,Mashhoon:1985cya,Berti:2005eb,Dolan:2009nk}. So, in principle, the $(3,\,3)$ and $(4,\,4)$ modes should allow us to probe masses that increase linearly with $\ell$ (and $m$). This effect is partially offset by the smaller amplitude of the higher modes and by cosmological redshift. If the radiated energy is large enough (or the noise power spectral density is low enough) that the signal is visible out to $z\gtrsim 1$, the observed frequency $f=f_{\rm s}/(1+z)$ of low-$\ell$ modes decreases by a significant factor with respect to the source-frame mode frequency $f_{\rm s}$, so low-mass BHs are ``redshifted back in band''  (as seen in Fig.~\ref{fig:fullHorizon}). At the moderate redshifts accessible to Advanced LIGO (ET), the ``eikonal limit enhancement'' effect for the $(3,\,3)$ and $(4,\,4)$ modes prevails (if only slightly) at masses of order $\sim 2\times 10^3~M_\odot$ ($\sim 2\times 10^4~M_\odot$, respectively), so these modes allow us to peer deeper into the IMBH regime.

%%%%%%%%%%%% Plot HorizonGW151914%%%%%%%%%%%%%%%%%%%
\begin{figure*}[t]
  \includegraphics[width=\columnwidth]{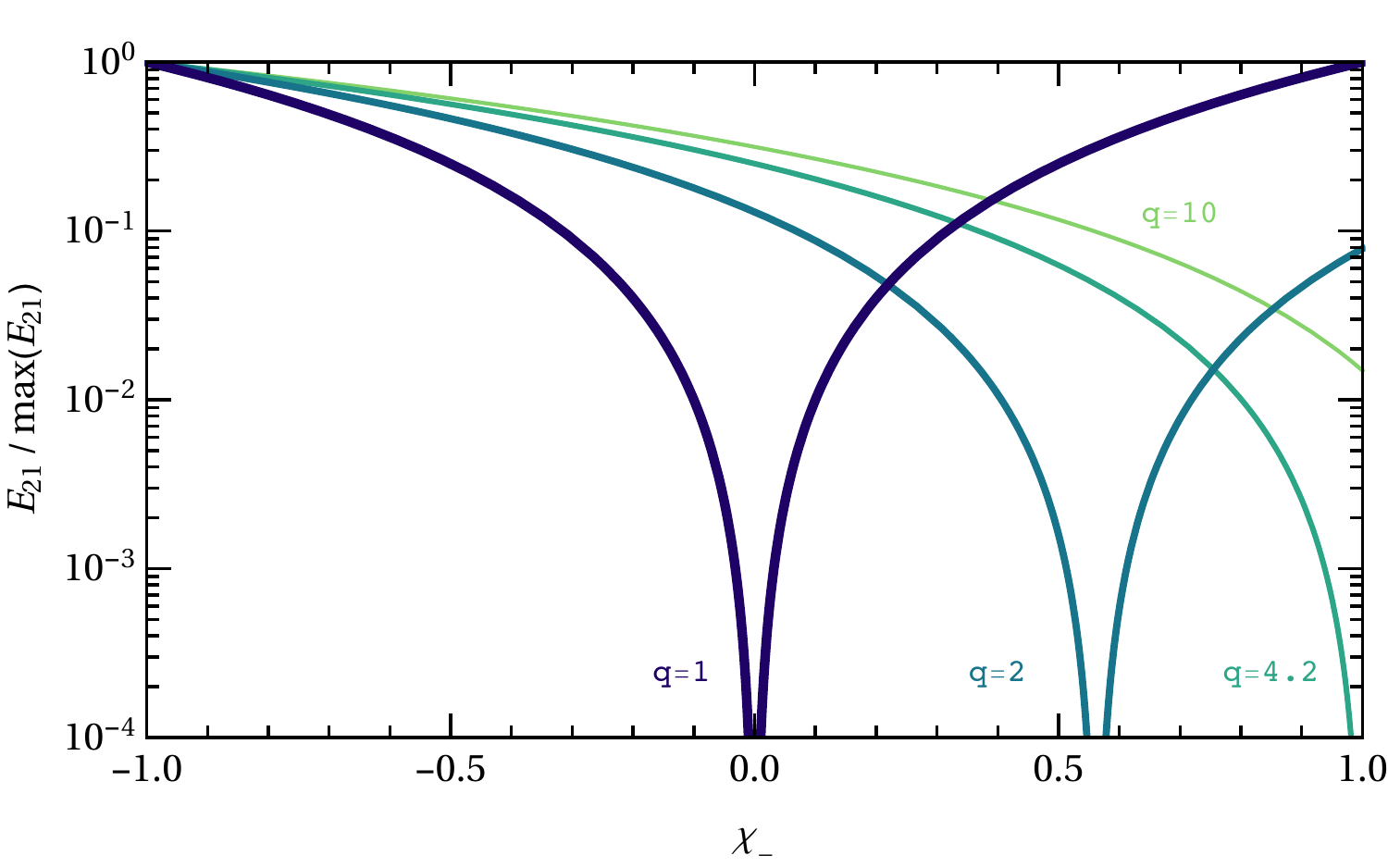}  
  \includegraphics[width=\columnwidth]{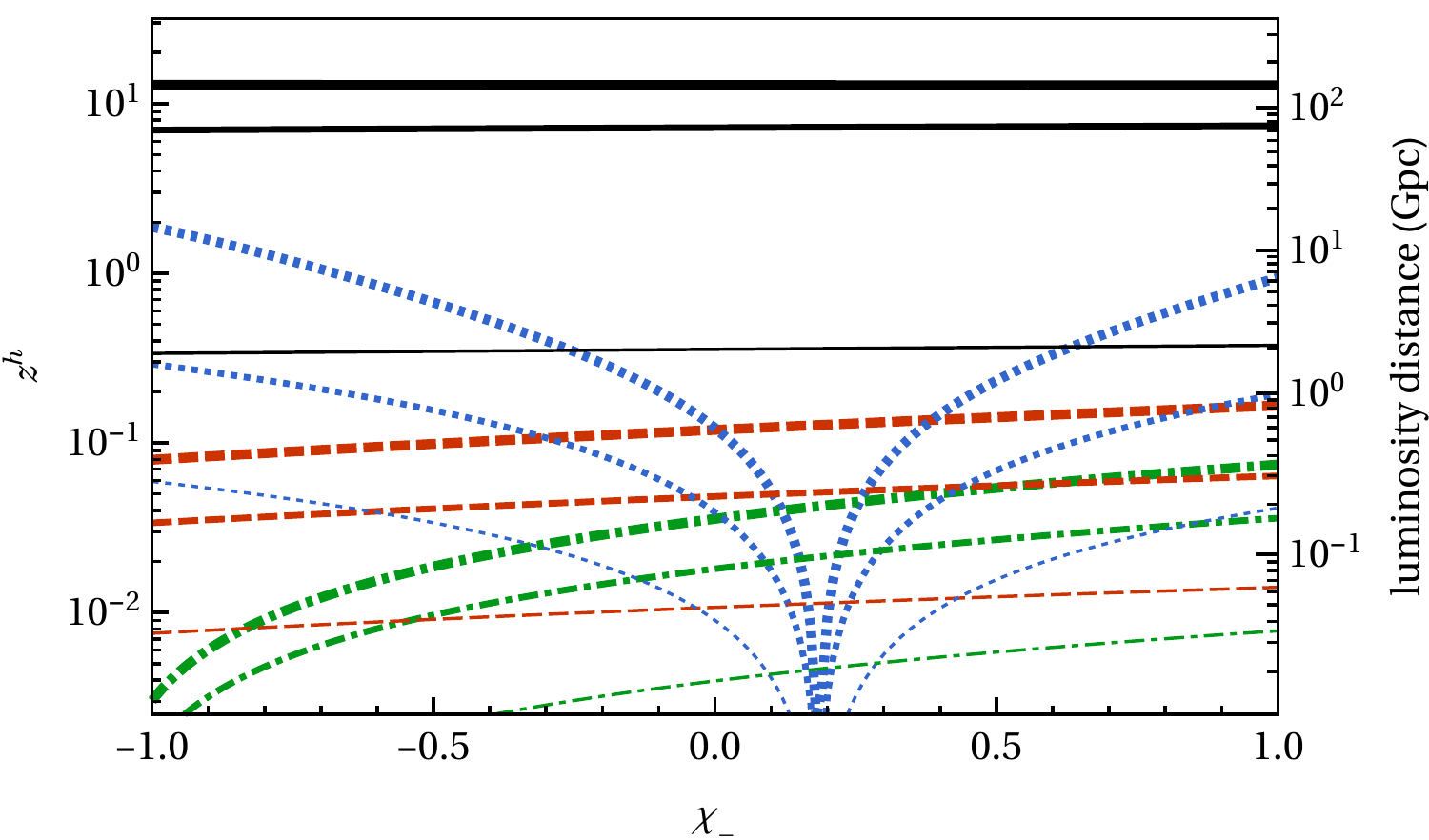}  
  \caption{Left: Energy in the $(2\,,1)$ mode normalized by its maximum value, which corresponds to $\chi_-=-1$~\cite{Baibhav:2017jhs}, as a function of $\chi_-$ for selected values of $q$. Right: Horizon redshift (left y-axis) and luminosity distance (right y-axis) for an optimally oriented GW150914-like binary as a function of $\chi_-$. Thick, medium and thin lines correspond to ET, Voyager and Advanced LIGO, respectively. Black, red, green and blue lines refer to the $(2,\,2)$, $(3,\,3)$, $(4,\,4)$ and $(2,\,1)$ modes, respectively.}
  \label{fig:HorizonGW151914}
\end{figure*}
%%%%%%%%%%%%%%%%%%%%%%%%%%%%%%%%%%%%%%%%%%%%%%%%%%%%%%%
%

%%%%%%%%%%%% Plot no of modes Best Worst ET%%%%%%%%%%%%%%%%%%%
\begin{figure}[t]
  \includegraphics[width=\columnwidth]{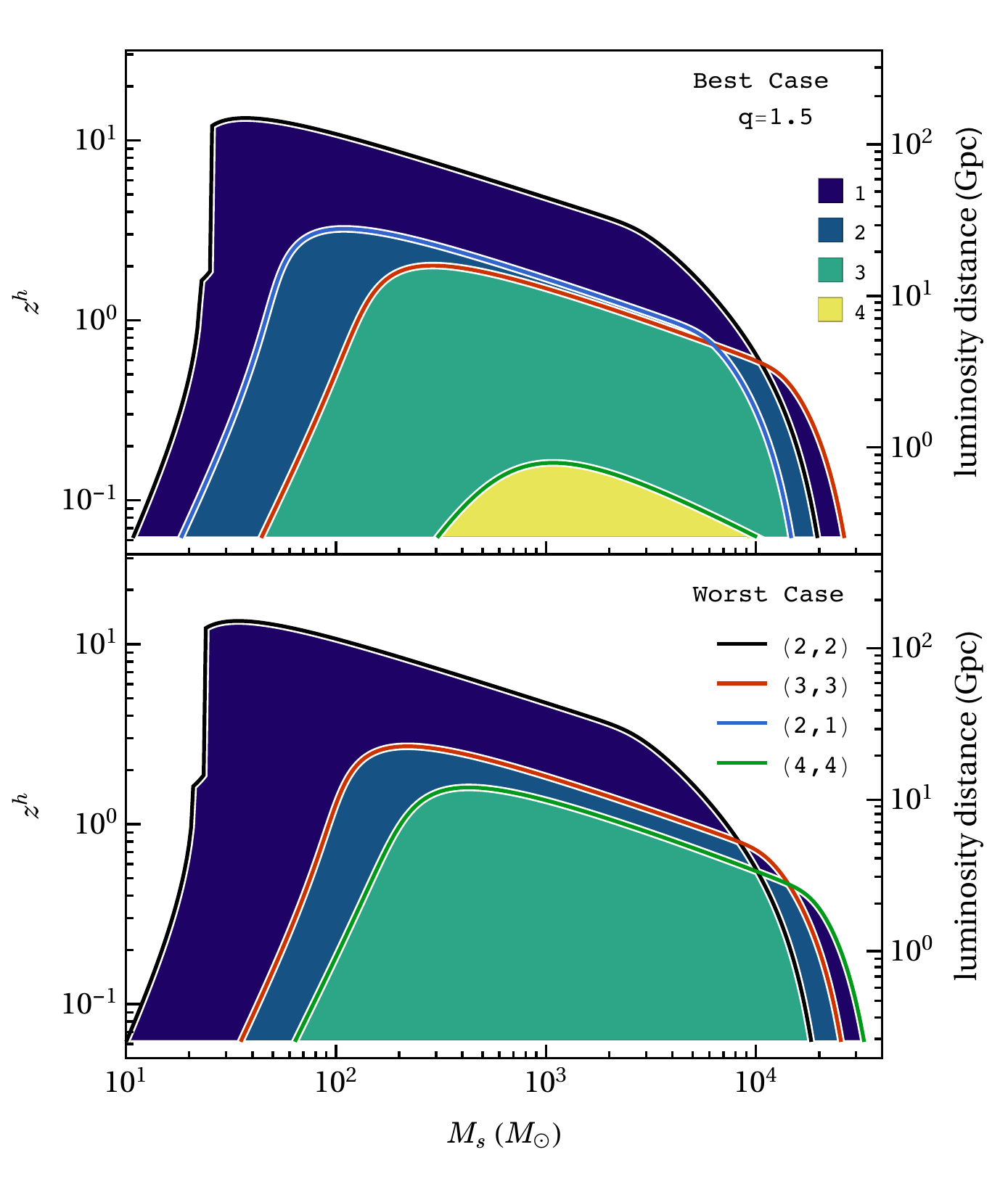}  
  \caption{ET horizon redshift and luminosity distance for the $(2\,,2)$, $(3\,,3)$, $(2\,,1)$ and $(4\,,4)$) modes as a function of source mass for BH mergers with mass ratio $q=1.5$. The best (worst) cases correspond to the value of $\chi_-$ that maximizes (minimizes) the energy radiated in the $(2\,,1)$ mode.} 
  \label{fig:noOfModesBestWorstET}
\end{figure}
%%%%%%%%%%%%%%%%%%%%%%%%%%%%%%%%%%%%%%%%%%%%%%%%%%%%%%%

In the case of LISA, by comparing the top and bottom panel we can see some important effects related to the cosmological redshift of observable masses and frequencies. The plot of the horizon redshift as a function of the {\em detector-frame} mass (bottom) traces quite closely the shape of the LISA noise power spectral density~\cite{Cornish:2018dyw}, including the characteristic ``bump'' due to galactic confusion noise (for which we assume four years of observation time). The detectability of IMBHs in the mass range between $\approx 10^4 M_\odot$ and $\approx 10^5 M_\odot$ depends on the LISA sensitivity at frequencies $\gtrsim 0.1$~Hz, which is uncertain. Similarly, the detection of the $(2,\,2)$ and $(2,\,1)$ modes for BHs of mass $M_{\rm s}\sim 10^9~M_\odot$ relies on understanding the noise power spectral density below $\sim10^{-5}$~Hz. To highlight these uncertainties, we use a dashed line to mark computed horizon redshifts that depend on the high- and low-frequency ends of the LISA noise power spectral density.

LISA will be sensitive to massive BH ringdowns in the $10^{5}$--$10^{9} M_{\odot}$ range out to very large redshifts. A remarkable feature of Fig.~\ref{fig:fullHorizon} is that LISA can detect ringdown modes from essentially all multipolar components computed by state-of-the-art numerical relativity simulations, up to $\ell=m=7$. In fact, even modes whose amplitude is comparable to numerical noise in current simulations -- such as the $(8,\,8)$ mode -- could be observable.
%Some lighter SMBHs ($\sim 10^5 M_{\odot}$) are more sensitive at higher redshifts, $z>1$  than they are at $z<1$. This can be explained as follows.\vv{Elaborate}
%Explain LISA redshift thing 
The LISA horizon redshift as a function of the source-frame mass (top panel) shows a characteristic ``turnover'' for IMBHs at $z\sim 1$: at such sizeable redshifts, ringdown signals at (source-frame) masses that would otherwise be unobservable are ``redshifted back'' in the LISA band and become observable. This is particularly interesting, because LISA ringdown signals can be used to probe the IMBH population at masses $M_{\rm s}\lesssim 10^5~M_\odot$ and redshifts $z\sim 10$, when mergers of these objects might have been common.

Figure~\ref{fig:fullHorizon} also shows that ground-based detectors are complementary to LISA in their potential to investigate the nature of IMBHs, being sensitive to multiple ringdown modes from IMBH remnants of source-frame mass $M_{\rm s}\lesssim 4\times 10^4~M_\odot$ at relatively small redshift. With Advanced LIGO, the $(2\,,2)$, $(3\,,3)$ or $(4\,,4)$ ringdown modes of IMBHs could be detected up to masses of $\sim1750 M_{\odot}$, $2780 M_{\odot}$ or $3760 M_{\odot}$, respectively. A third-generation detector like ET can observe IMBHs with masses up to an order of magnitude larger than this.

%%%%%%%%%%%% Plot no of modes Best Worst LISA%%%%%%%%%%%%%%%%%%%
\begin{figure*}[thp]
  \includegraphics[width=\columnwidth]{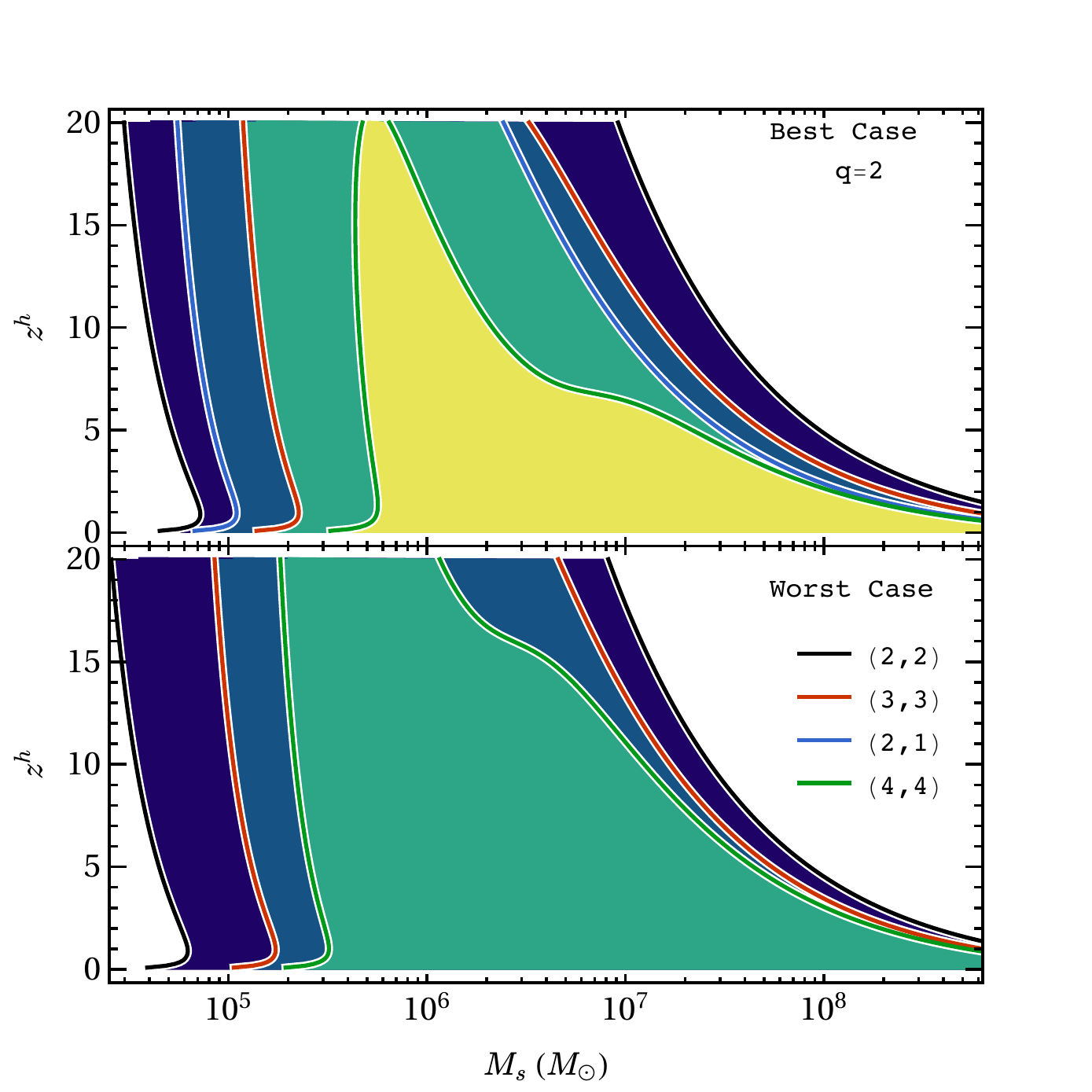}  
   \includegraphics[width=\columnwidth]{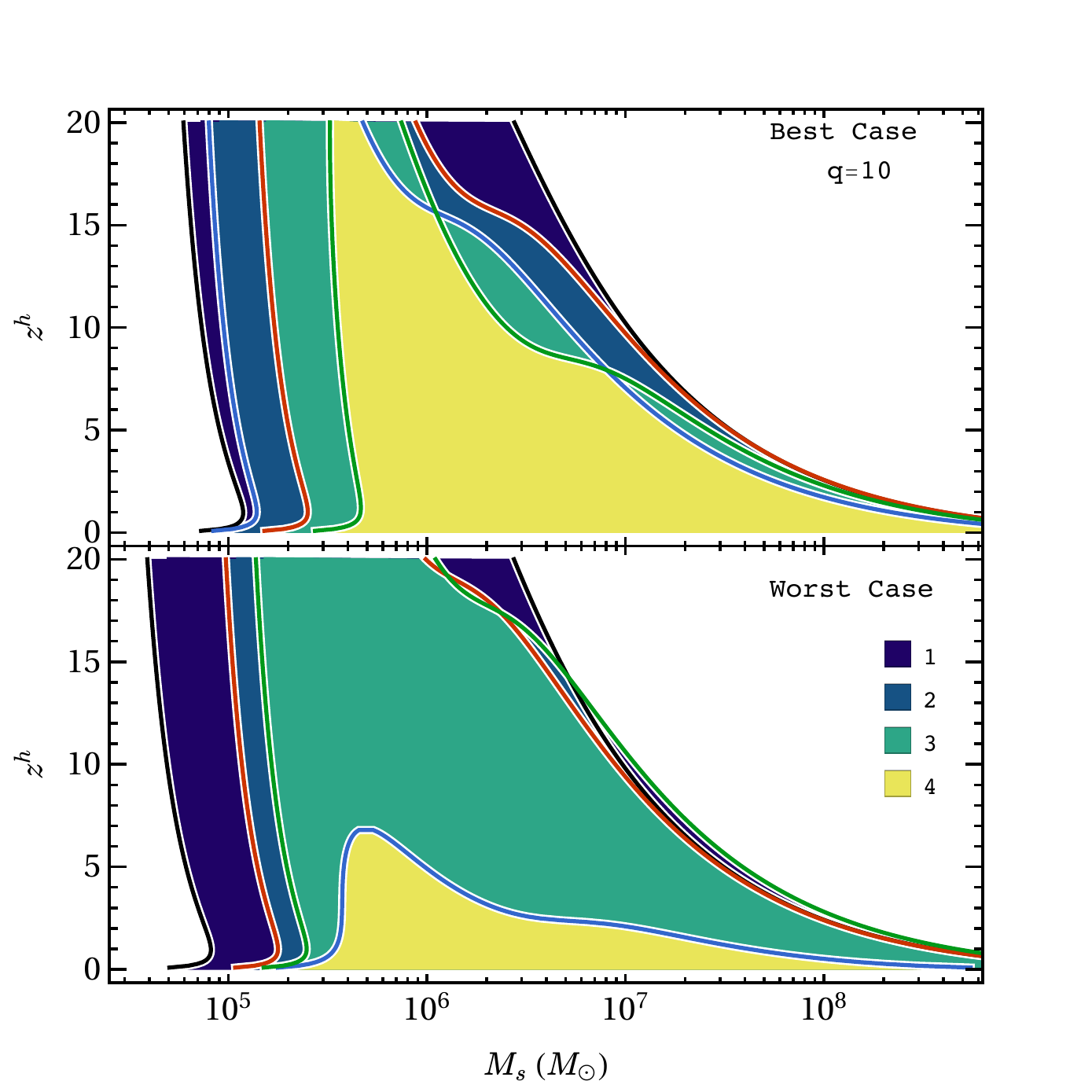}  
   \caption{LISA horizon redshift and luminosity distance for the $(2\,,2)$, $(3\,,3)$, $(2\,,1)$ and $(4\,,4)$) modes as a function of source mass for BH mergers with mass ratio $q=2$ (left panels) and $q=10$ (right panels). Estimates for the best/worst case were found by choosing the value of $\chi_-$ that maximizes/minimizes the energy radiated in the $(2\,,1)$ mode.} 
  \label{fig:noOfModesBestWorstLISA}
\end{figure*}
%%%%%%%%%%%%%%%%%%%%%%%%%%%%%%%%%%%%%%%%%%%%%%%%%%%%%%%

\subsection{Effect of spins on multi-mode ringdown observations}

How do spins affect these ringdown horizon estimates?
Let us first consider, for concreteness, the BH mergers observed during the O1 and O2 runs. Only the first event (GW150914) had a marginally detectable ringdown signal, and most binaries had a measured (symmetric) effective spin parameter $\chi_+$ compatible with zero. The data do not place strong constraints on the magnitude of the individual spins (and consequently, on the asymmetric effective spin parameter $\chi_-$).

In \cite{Baibhav:2017jhs} we estimated the energy radiated in each multipole by fitting numerical relativity simulations. Confirming earlier conclusions~\cite{Berti:2007nw,Kamaretsos:2012bs}, we found that the excitation of the $\ell=m=2$, $3$, $4$ modes depend weakly on the spins for comparable mass ratios, while the $(2\,,1)$ mode strongly depends on the spins (at leading order) through the poorly constrained parameter $\chi_-$. In the left panel of Fig.~\ref{fig:HorizonGW151914} we plot this dependence for selected values of the mass ratio $q$. For $q\lesssim 4.2$ the energy radiated in the $(2\,,1)$ mode vanishes, and therefore the mode becomes unobservable, at some finite value of $\chi_-<1$ which is well approximated by (using fits from \cite{Baibhav:2017jhs})
\be\label{eq:chispin21}
\chi_-\simeq \frac{q-1}{q+1} \left[1.49+\frac{0.9   q}{(q+1)^2}\right]\,.
\ee

Values of $q$ and $\chi_-$ such that the energy in the $(2,\,1)$ mode vanishes are (in this sense) worst-case scenarios for the observation of multiple modes. The best-case scenario is the one that yields the maximum horizon redshift for the $(2,1)$ mode. As we see from the left panel of Fig.~\ref{fig:HorizonGW151914}, this corresponds to $\chi_-=-1$ for all values of $q$.
To quantify how uncertainties in $\chi_-$ would impact multi-mode spectroscopy, in the right panel of Fig.~\ref{fig:HorizonGW151914} we plot the horizon redshift for the dominant multipoles of a GW150914-like binary with total source-frame mass $m_{1 {\rm s}}+m_{2 {\rm s}}=65 M_\odot$, mass ratio $q\simeq 1.24$, and $\chi_+\simeq 0$.
The horizon redshift and luminosity distance for each mode increases as the detectors become more sensitive, but it is always a mildly varying function of $\chi_-$ for the $(2,\,2)$, $(3,\,3)$ and $(4,\,4)$ modes. However the amplitude of the $(2,\,1)$ mode drops to zero (and the mode becomes unobservable) when $\chi_-\simeq 0.18$.

In Fig.~\ref{fig:noOfModesBestWorstET} we show the best- and worst-case horizon redshifts for an optimally oriented binary as observed by ET. For concreteness we focus on a mass ratio $q=1.5$, close to the mean measured mass ratio for BH binaries detected so far by the LIGO/Virgo collaboration.\footnote{As pointed out in previous work, numerical merger simulations of spinning binaries with comparable masses ($q\simeq 1$) yield unreliable estimates for the ringdown energy in the $(4\,,4)$ mode~\cite{Baibhav:2017jhs}.} For $q=1.5$, the worst-case scenario where the $(2,\,1)$ mode is undetectable corresponds to $\chi_-=0.34$. The top panel shows that, in the best-case scenario, ET could observe as many as three (four) modes out to $z\sim 2.27$ ($z\sim 0.18$) for mergers with total source mass $M_{\rm s}\sim 200~M_\odot$. Even in the worst-case scenario, we could observe three modes out to $z\sim 1$ if IMBH mergers of total mass $M_{\rm s}\sim 600~M_\odot$ occur in the local Universe.

In Fig.~\ref{fig:noOfModesBestWorstLISA} we show the best- and worst-case horizon redshifts for optimally oriented binaries with $q=2$ (left) and $q=10$ (right) as observed by LISA. These mass ratios were chosen to bracket the typical range of mass ratios expected from astrophysical models of BH formation (see e.g. Fig.~3 of \cite{Sesana:2010wy}). For $q=2$, the $(2,\,1)$ mode is undetectable when $\chi_-=0.56$; for $q=10$, it is least excited when $\chi_-=1$. It is truly remarkable that LISA can observe the four dominant modes (and in fact, also many of the subdominant modes, not shown in this plot) out to $z> 20$ for mergers with source mass $M_{\rm s}\sim 5\times 10^5~M_\odot$. Even in the worst-case scenario, depending on the masses of the merging BHs, LISA could see the four dominant modes out to redshift $z\gtrsim 5$.

%%%%%%%%%%%% Plot projection CDF%%%%%%%%%%%%%%%%%%%
\begin{figure*}[htp] 
  \includegraphics[width=\columnwidth]{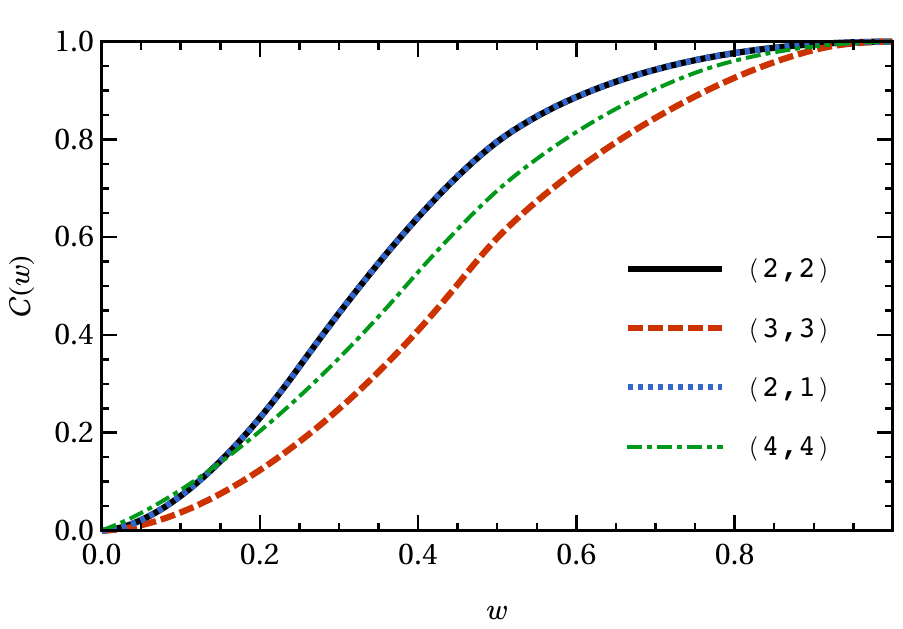}  
  \includegraphics[width=\columnwidth]{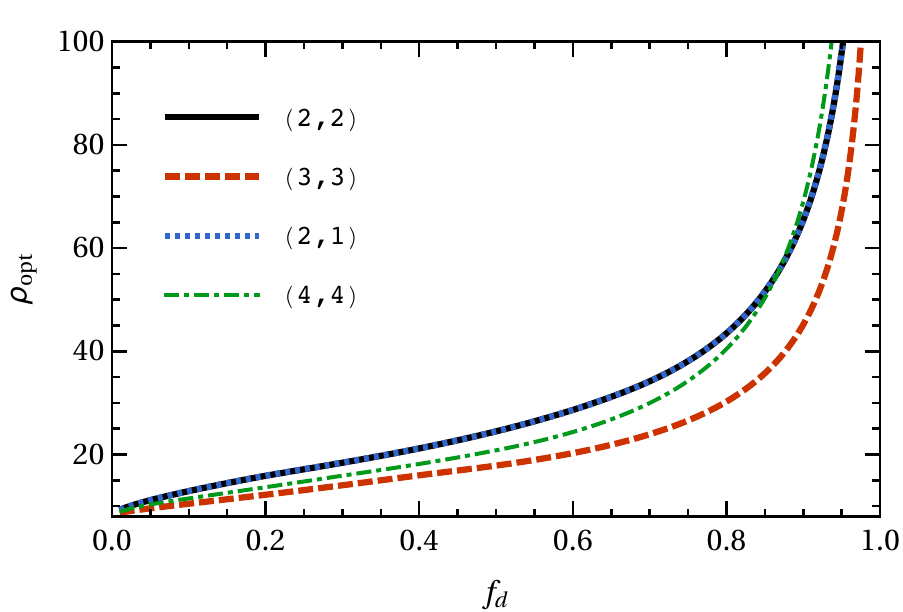}  
  \caption{Left: Cumulative distribution function ${\mathcal C}(w)$ for the projection parameter $w$. Right: Optimal SNR required by a ringdown mode to be detected with probability $f_{\rm d}$.}
\label{fig:projectionCDF}
\end{figure*}
%%%%%%%%%%%%%%%%%%%%%%%%%%%%%%%%%%%%%%%%%%%%%%%%%%%%%%%

%%%%%%%%%%%% Plot response redshift ET adLIGO%%%%%%%%%%%%%%%%%%%
\begin{figure}[htp]
  \includegraphics[width=\columnwidth]{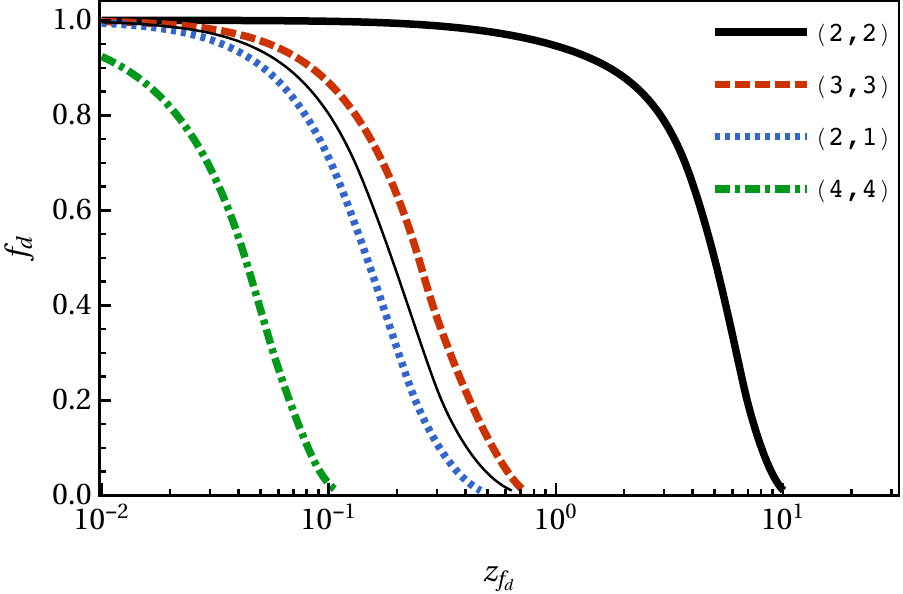}  
  \caption{Response redshift $z_{f_{\rm d}}$ at which a $100 M_\odot$ nonspinning binary with $q=1.5$ could be detected with probability $f_{\rm d}$ by ET (thick lines) or Advanced LIGO (thin line).}
%    \vv{I could have chosen a different mass but then LIGO 22 and ET 33 lines overlap. Also, we can say that $100 M_\odot$ is maximum possible through merger of stellar mass BHs.}
  \label{fig:responseRedshiftETLigo}
\end{figure}
%%%%%%%%%%%%%%%%%%%%%%%%%%%%%%%%%%%%%%%%%%%%%%%%%%%%%%%

%%%%%%%%%%%% Plot response redshift%%%%%%%%%%%%%%%%%%%
\begin{figure*}[htp]
  \includegraphics[width=0.9\textwidth]{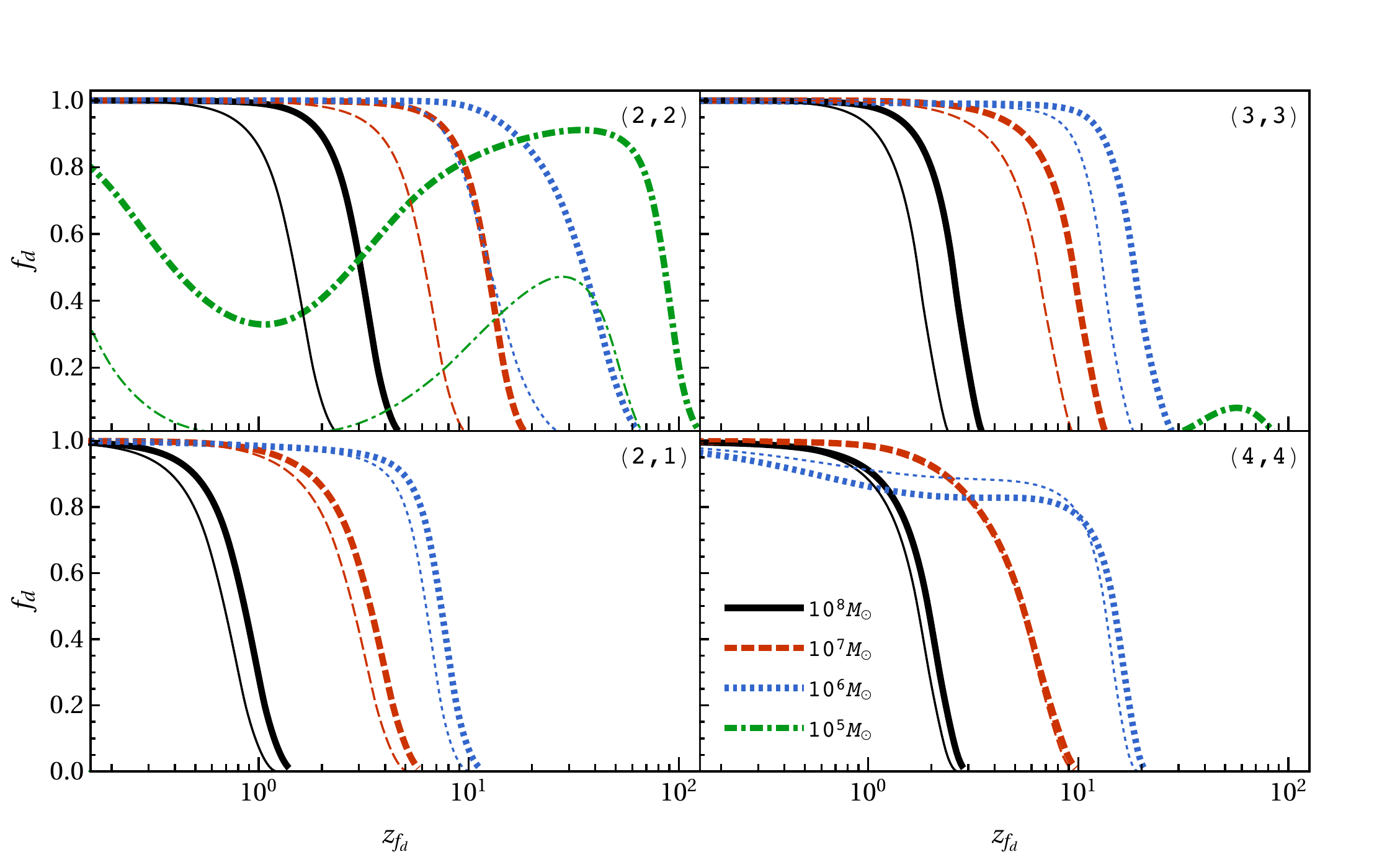}  
  \caption{Response redshift $z_{f_{\rm d}}$ at which nonspinning binaries of selected source-frame masses with $q=2$ (thick lines) and $q=10$ (thin lines) could be detected with probability $f_{\rm d}$ by LISA.}
  \label{fig:responseRedshiftLISA}
\end{figure*}
%%%%%%%%%%%%%%%%%%%%%%%%%%%%%%%%%%%%%%%%%%%%%%%%%%%%%%%

\subsection{Response redshift and detectability fraction}

The horizon estimates computed so far assume optimal source orientation and sky location. In general, the SNR of observed events depends on the source position and orientation. The sky sensitivity of the detector (as encoded in $w_{\ell m}$) can affect the detectability of individual modes (see~\cite{Finn:1992wt,Chernoff:1993th,Belczynski:2014iua,OShaughnessy:2010igj,OShaughnessy:2009szr,Finn:1995ah,Dominik:2014yma} for a discussion of this issue in the context of inspiral, and \cite{Chen:2017wpg} for a nice overview of the nomenclature and conventions used in the gravitational-wave literature).

The cumulative distribution function for the ``projection function'' $w_{\ell m}$ is independent of the intrinsic properties of the source. It can be generated through a Monte Carlo over sky location and source orientation. Because of the $e^{{\rm i} m \phi}$ dependence of the spin-weighted spherical harmonics, the cumulative distribution function depends only on $\ell$: for example, ${\mathcal C}(w)$ for the $(2\,,1)$ mode coincides with the well-known cumulative distribution function for the $(2\,,2)$ mode of the inspiral and ringdown. The cumulative distribution functions for the dominant modes are shown in the left panel of Fig.~\ref{fig:projectionCDF}. Under the assumption that a binary is detectable when $w_{\ell m}>8/\rho_{\rm opt}$, from the cumulative distribution function we can also compute the fraction of detectable binaries
\be
f_{\rm d}=1-{\mathcal C}(8/\rho_{\rm opt})\,,
\ee
Just like ${\mathcal C}(w)$, $f_{\rm d}$ depends on $\ell$ but not on $m$. From the right panel of Fig.~\ref{fig:projectionCDF} we infer that the optimal SNR $\rho_{\rm opt}$ required to detect the $(2,\,2)$, ($3,\,3$) and $(4,\,4)$ mode with $50\%$ ($95\%$) probability is $24.5$ ($98$), $17.8$ ($67.3$) and $20.8$ ($121.2$), respectively.  For the $(2,\,2)$ mode, detection probabilities $f_{\rm d}=0.36$, $0.63$, $0.84$ and $0.95$ correspond to $\rho_{\rm opt}=20$, $30$, $50$ and $100$, respectively.

Redshift horizons computed by setting $\rho_{\rm opt}=8$ do not give information about the probability of detecting binaries with suboptimal orientations. For this purpose it is useful to introduce the ``response redshift'' $z_{f_{\rm d}}$ (see e.g.~\cite{Chen:2017wpg}), defined as the redshift at which a binary could be detected with probability $f_{\rm d}$. In Fig.~\ref{fig:responseRedshiftETLigo} we plot $f_{\rm d}$  as a function of $z_{f_{\rm d}}$ for the $(2,\,2)$ mode in the case of Advanced LIGO (thin line), and for the four dominant modes in the case of ET (thick lines). All plots refer to a $100 M_\odot$ nonspinning binary with $q=1.5$.
By definition, the detection probability $f_{\rm d}$ drops to zero at the horizon redshift, where $\rho_{\rm opt}=8$. For ET, the detection probability $f_{\rm d}$ is $90\%$ ($50\%$) at $z=1.72$ ($4.99$), $0.08$ ($0.25$), $0.05$ ($0.15$) and $0.01$ ($0.04$) for the $(2,\,2)$, $(3,\,3)$, $(2,\,1)$ and $(4,\,4)$ modes, respectively. For Advanced LIGO, the detection probability $f_{\rm d}$ is $90\%$ ($50\%$) at $z=0.07$ ($0.19$) for the $(2,\,2)$ mode. Note that the redshift at which the detection probability $f_{\rm d}=0.5$ is a useful indicator of the distance at which binary ringdown is observable, because it corresponds to the {\em median redshift} at which the given mode would be visible, independently of astrophysical assumptions on the intrinsic merger rates~\cite{Chen:2017wpg}.

Figure~\ref{fig:responseRedshiftLISA} shows similar results for LISA observations of BHs of total source-frame mass $10^5M_{\odot}$, $10^6M_{\odot}$, $10^7M_{\odot}$ and $10^8M_{\odot}$ with either $q=2$ (thick lines) or $q=10$ (thin lines). Once again, by definition $f_{\rm d}=0$ at the horizon redshift, where $\rho_{\rm opt}=8$. The behavior of these probability distributions when the source mass is $10^6M_{\odot}$, $10^7M_{\odot}$ and $10^8M_{\odot}$ is very similar to the results shown in Fig.~\ref{fig:responseRedshiftETLigo}. However, the probability distribution for binaries of mass $10^5M_{\odot}$ shows an interesting bimodal distribution. This bimodality is related to the characteristic ``turnover'' for IMBHs at $z\sim 1$ that we observed in Fig.~\ref{fig:fullHorizon}: when $z\gtrsim 1$, ringdown signals at (source-frame) masses that would otherwise have been unobservable are ``redshifted back'' in the LISA band and become observable. The high-redshift peak observed in the dash-dotted (green) probability distributions in the top panels of Fig.~\ref{fig:responseRedshiftLISA} has very interesting implications for IMBH mergers at high redshifts: LISA ringdown signals can be used to probe the populations of IMBHs with mass $M_{\rm s}\lesssim 10^5~M_\odot$ and redshifts $z\gtrsim 10$. This could be a unique way to shed light on the formation and merger of primordial BHs and of IMBH seeds produced in more conventional scenarios, such as the relativistic collapse of massive Pop III stars or the direct collapse of a supermassive protostar in a metal-free dark matter halo (see e.g.~\cite{Barack:2018yly}).

Tables \ref{tab:Ground} and \ref{tab:LISA} complement and extend the results in Figures~\ref{fig:responseRedshiftETLigo} and \ref{fig:responseRedshiftLISA}.
In these Tables we list the horizon redshift for an optimally oriented binary and (in parentheses) the redshift corresponding to the median value of sky sensitivity $\Omega_{\ell m}$, i.e. the response redshift at which
$f_{\rm d}=0.5$.

Table~\ref{tab:Ground} lists these quantities for nonspinning binaries with $q=1.5$, selected values of the remnant source-frame BH mass $M_{\rm s}$, and three ground-based detectors (Advanced LIGO, Voyager and ET). Advanced LIGO cannot observe subdominant modes from the merger of stellar-mass BH binaries, but it could observe the $(3,\,3)$ mode out to redshifts $z\gtrsim 0.2$ for IMBH mergers with $M_{\rm s}\sim 500~M_\odot$. The horizon and median redshift decrease when $M_{\rm s}\sim 10^3~M_\odot$ for all ground-based detectors, but the better low-frequency sensitivity of ET makes it possible to observe multiple ringdown modes out to $z\sim 1$ from BH remnants as massive as $M_{\rm s}\sim 5000~M_\odot$. Note also that the ET horizon redshift for a $M_{\rm s}\sim 50~M_\odot$ binary is $z^h\sim 13$, but the median redshift is much lower ($z_{0.5}\sim 3$): these findings are compatible with previous rate calculations based on population synthesis models (see e.g. Fig.~3 of \cite{Berti:2016lat}, and Figs.~13 and 15 of \cite{Belczynski:2016ieo}).

Table~\ref{tab:LISA} lists these quantities for LISA observations of BH binary mergers with $q=2$ and $q=10$. The observed trends are easily explained by considering that the relative excitation of subdominant modes is higher, but the total energy radiated (and therefore the horizon redshift) decrease when $q$ gets larger~\cite{Berti:2007fi}. LISA has the incredible potential to measure multiple ringdown modes in a wide range of masses and mass ratios out to cosmological redshifts. Interestingly, LISA can observe multiple ringdown modes from very massive binary mergers (say, $10^8+10^9~M_\odot$), as long as the merger rates are high enough in the local Universe ($z\lesssim 0.5$): see e.g.~\cite{Katz:2018dgn} for a recent, detailed investigation of how LISA design choices would affect this science. We plan to explore this possibility using astrophysical BH formation models in future work.

%%%%%%%%%%%%%%%%%% LIGO
\begin{table*}
  \caption{Horizon redshift out to which a given mode can be detected with ground-based detectors for nonspinning binaries with $q=1.5$ and selected values of the remnant BH mass in the source frame, $M_{\rm s}$. The computed horizon redshifts assume either optimal orientation ($z^h$) or $f_{\rm d}=0.5$ ($z_{0.5}$, in parenthesis): see text for details.}
\label{tab:Ground}
\setlength\tabcolsep{9 pt}
\begin{tabular}{LCCCC}
\hline
\hline
M_{\rm s}(M_\odot)& (2,\,2) & (3,\,3)& (2,\,1)& (4,\,4)\\
\hline
& z^h\,(z_{0.5}) & z^h\,(z_{0.5}) & z^h\,(z_{0.5}) & z^h\,(z_{0.5}) \\
\hline
\multicolumn{5}{C}{\text{Advanced LIGO}}\\
\hline
  50 & 0.21\,(0.06) & 0.01\,(0.01) & 0.01\,(0.00) & 0.00\,(0.00) \\
 100 & 0.80\,(0.19) & 0.04\,(0.02) & 0.03\,(0.01) & 0.01\,(0.00) \\
 500 & 1.25\,(0.60) & 0.32\,(0.15) & 0.12\,(0.04) & 0.11\,(0.04) \\
 1000 & 0.65\,(0.40) & 0.30\,(0.15) & 0.08\,(0.03) & 0.16\,(0.07) \\
\hline
\hline
\multicolumn{5}{C}{\text{Voyager}}\\
\hline
 50 & 2.53\,(0.32) & 0.06\,(0.03) & 0.05\,(0.02) & 0.01\,(0.01) \\
 100 & 7.15\,(1.41) & 0.18\,(0.08) & 0.14\,(0.04) & 0.05\,(0.02) \\
 500 & 2.30\,(1.61) & 1.45\,(0.73) & 0.54\,(0.23) & 0.54\,(0.18) \\
 1000 & 0.99\,(0.75) & 0.94\,(0.62) & 0.26\,(0.12) & 0.73\,(0.33) \\
\hline
\hline
\multicolumn{5}{C}{\text{Einstein Telescope}}\\
\hline
 50 & 13.03\,(3.03) & 0.13\,(0.05) & 0.14\,(0.04) & 0.02\,(0.01) \\
 100 & 11.04\,(4.99) & 0.85\,(0.25) & 0.56\,(0.15) & 0.12\,(0.04) \\
 500 & 6.33\,(3.24) & 2.26\,(1.28) & 0.89\,(0.39) & 1.28\,(0.54) \\
 1000 & 4.89\,(2.51) & 1.81\,(1.07) & 0.70\,(0.31) & 1.16\,(0.57) \\
 5000 & 1.86\,(1.31) & 0.97\,(0.56) & 0.34\,(0.14) & 0.63\,(0.32) \\
\hline
\hline
\end{tabular}
\end{table*}
%%%%%%%%%%%%%%%%%%%%%%%%%%%%%%%%%

%%%%%%%%%%%%%%%LISA
\begin{table*}
  \caption{Horizon redshift out to which a given mode can be detected with LISA for selected nonspinning binaries with $q=2$ (top) and $q=10$ (bottom). The computed horizon redshifts assume either optimal orientation ($z^h$) or $f_{\rm d}=0.5$ ($z_{0.5}$, in parenthesis): see text for details.}
\label{tab:LISA}
\setlength\tabcolsep{9 pt}
\begin{tabular}{LCCCC}
\hline
\hline
(m_{1 {\rm s}}+m_{2 {\rm s}})(M_\odot)& (2,\,2) & (3,\,3)& (2,\,1)& (4,\,4)\\
\hline
& z^h\,(z_{0.5}) & z^h\,(z_{0.5}) & z^h\,(z_{0.5}) & z^h\,(z_{0.5}) \\
\hline
\multicolumn{5}{C}{q=2}\\
\hline
 2\times10^6+10^6 & 38.78\,(23.48) & 24.04\,(13.51) & 9.95\,(4.3) & 13.33\,(7.61) \\
 2\times10^7+10^7 & 10.18\,(6.45) & 7.46\,(5.29) & 3.29\,(1.95) & 6.1\,(3.84) \\
 2\times10^8+10^8 & 2.43\,(1.46) & 1.75\,(1.19) & 0.65\,(0.33) & 1.43\,(0.89) \\
 2\times10^9+10^9 & 0.42\,(0.20) & 0.26\,(0.14) & 0.05\,(0.02) & 0.19\,(0.09) \\
\hline
\hline
\multicolumn{5}{C}{q=10}\\
\hline
 10^5+10^6 & 28.67\,(12.24) & 18.58\,(12.35) & 9.82\,(5.88) & 18.24\,(12.61) \\
 10^6+10^7 & 9.57\,(5.87) & 9.22\,(6.17) & 4.92\,(2.71) & 9.3\,(5.23) \\
 10^7+10^8 & 2.32\,(1.39) & 2.33\,(1.61) & 1.14\,(0.64) & 2.5\,(1.61) \\
 10^8+10^9 & 0.40\,(0.18) & 0.40\,(0.24) & 0.14\,(0.05) & 0.44\,(0.24) \\
\hline
\hline
\end{tabular}
\end{table*}
%%%%%%%%%%%%%%%%%%%%%%%%%%%%%

%%%%%%%%%%%%%%%%%%%%%%%%%%%%%%%%%%%%%%%%%%%%%%%%%
%%%%%%%%%%%%%%%%%%%%%%%%%%%%%%%%%%%%%%%%%%%%%%%%
\section{Conclusions and outlook}
\label{sec:conclusions}
%%%%%%%%%%%%%%%%%%%%%%%%%%%%%%%%%%%%%%%%%%%%%%%%%%%%%
%%%%%%%%%%%%%%%%%%%%%%%%%%%%%%%%%%%%%%%%%%%%%%%%%%%%%

Atomic spectroscopy is a standard tool in modern astronomy. As gravitational wave detectors improve in sensitivity, it seems reasonable to expect that gravitational spectroscopy will similarly become a standard tool to identify merger remnants as the Kerr BHs predicted by general relativity, unless nature has some surprise in store.

In this paper we investigated the potential of future detectors to detect multiple gravitational spectral lines. We computed the horizon and median redshifts at which the dominant modes of the radiation can be detected by Advanced LIGO, third-generation ground-based detectors such as ET, and LISA.

We found that Advanced LIGO cannot observe subdominant modes from the merger of stellar-mass BH binaries, but it could observe the $(3,\,3)$ mode out to redshifts $z^h\gtrsim 0.2$ for IMBH mergers with $M_{\rm s}\sim 500~M_\odot$. Horizon redshifts decrease when $M_{\rm s}\sim 10^3~M_\odot$ for all ground-based detectors, but the better low-frequency sensitivity of ET makes it possible to observe multiple ringdown modes out to $z^h\sim 1$ from BH remnants as massive as $M_{\rm s}\sim 5000~M_\odot$. The ET horizon redshift for a $M_{\rm s}\sim 50~M_\odot$ binary can be very large ($z^h\sim 13$), but the median redshift is much lower ($z_{0.5}\sim 3$). In contrast, BH binary mergers in the LISA band could be used to measure multiple ringdown modes in a wide range of masses and mass ratios out to cosmological redshifts. In fact, LISA can detect ringdown modes from all multipolar components computed so far in state-of-the-art numerical relativity simulations. Even modes whose amplitude is comparable to numerical noise in current simulations -- such as the $(8,\,8)$ mode -- could be observable.  Cosmological redshift produces a characteristic ``turnover'' in the LISA horizon redshift for IMBHs at $z\sim 1$ (Fig.~\ref{fig:fullHorizon}) and a bimodal distribution in the detection probability (Fig.~\ref{fig:responseRedshiftLISA}): large-$z$ ringdown signals at (source-frame) masses that would otherwise be unobservable are ``redshifted back'' in the LISA band and become observable. Therefore LISA observations of the merger/ringdown phase can be used to probe the IMBH population at masses $M_{\rm s}\lesssim 10^5~M_\odot$ and redshifts $z\sim 10$, when such mergers might have been common. LISA can also detect multiple ringdown modes from BH binary mergers of mass $\gtrsim 10^8~M_\odot$ at $z\lesssim 0.5$, as long as the merger rates are large enough in the local Universe. As pointed out in~\cite{Katz:2018dgn}, these science goals should be taken into account in design studies of the LISA sensitivity at low and high frequencies.

Our work can and should be improved in many ways. To quantify detectability we used fits from~\cite{Baibhav:2017jhs}, which are based on the ``energy maximized orthogonal projection'' criterion and more conservative than the estimates of Ref.~\cite{Berti:2016lat}, where we used the detection-oriented ``matched filtering'' fits from~\cite{Berti:2007fi}. Several different ways to quantify ringdown excitation from numerical simulations have been proposed over the years~\cite{Berti:2007fi,Berti:2007zu,Kamaretsos:2011um,Kamaretsos:2012bs,London:2014cma,Bhagwat:2017tkm,London:2018gaq}. In general these estimates lead to slightly different predictions for the horizon redshift. This dependence should be investigated. We plan to revise our fits -- especially for the spin dependence of subdominant modes, such as the $(4,\,4)$ mode -- as soon as updates to the SXS public waveform catalog described in~\cite{Mroue:2013xna} become available. but the broad qualitative conclusions of our work should remain valid. 

One of the main conclusions of our work is that LISA may allow us to see so many ringdown modes that systematic errors in numerical relativity simulations may be comparable to statistical errors. We hope that this consideration will motivate further studies of ringdown excitation and the development of more accurate numerical relativity simulations of BH mergers.

Of course, deviations from general relativity may drastically modify the ringdown spectrum~\cite{Berti:2015itd,Glampedakis:2017dvb,Tattersall:2017erk,Berti:2018vdi,Tattersall:2018map,Tattersall:2018nve}, and possibly even make subdominant modes undetectable. This possibility would invalidate our analysis, which assumes that general relativity is the correct theory of gravity. It would be very exciting if our study of subdominant ringdown modes were to prove wrong or irrelevant for this reason. 

\vspace{-0.1cm}
\acknowledgments

We thank Leo Stein for useful discussion. V.B. and E.B. are supported by NSF Grants No. PHY-1841464 and AST-1841358.

\bibliography{GLRT}

\end{document}